\documentclass{sig-alternate}

\begin{document}
%
% --- Author Metadata here ---
\conferenceinfo{arXiv.org}{'2015'}
%\CopyrightYear{2007} % Allows default copyright year (20XX) to be over-ridden - IF NEED BE.
%\crdata{0-12345-67-8/90/01}  % Allows default copyright data (0-89791-88-6/97/05) to be over-ridden - IF NEED BE.
% --- End of Author Metadata ---

\title{Efficient FFT mapping on GPU for radar processing application: modeling and implementation}
%\subtitle{[Extended Abstract]


\numberofauthors{5} %  in this sample file, there are a *total*
\author{
% 1st. author
 Mohamed Amine Bergach\\
       \affaddr{Kontron}\\
       \affaddr{Toulon, France}\\
       \email{mohamed.bergach@kontron.com}
% 2nd. author
\and
\alignauthor
Emilien Kofman\\
      \affaddr{Univ. Nice Sophia Antipolis}\\
      \affaddr{Nice, France}\\
      \email{emilien.kofman@inria.fr}
% 3rd. author
\alignauthor Robert de Simone\\
      \affaddr{INRIA}\\
      \affaddr{Nice, France}\\
      \email{robert.de\_simone@inria.fr}
  % use '\and' if you need 'another row' of author names
% 4th. author
\and
\alignauthor 
Serge Tissot\\
      \affaddr{Kontron}\\
      \affaddr{Toulon, France}\\
      \email{serge.tissot@kontron.com}
% 5th. author
\alignauthor 
Michel Syska\\
      \affaddr{Univ. Nice Sophia Antipolis}\\
      \affaddr{Nice, France}\\
      \email{michel.syska@inria.fr}
}
% There's nothing stopping you putting the seventh, eighth, etc.
% author on the opening page (as the 'third row') but we ask,
% for aesthetic reasons that you place these 'additional authors'
% in the \additional authors block, viz.

% Just remember to make sure that the TOTAL number of authors
% is the number that will appear on the first page PLUS the
% number that will appear in the \additionalauthors section.

\maketitle
\begin{abstract}
General-purpose multiprocessors (as, in our case, Intel IvyBridge and Intel Haswell) increasingly add GPU computing power to the former multicore architectures. When used for embedded applications (for us, Synthetic aperture radar) with intensive signal processing requirements, they must constantly compute convolution algorithms, such as the famous Fast Fourier Transform. Due to its "fractal" nature (the typical butterfly shape, with larger FFTs defined as combination of smaller ones with auxiliary data array transpose functions), one can hope to compute analytically the size of the largest FFT that can be performed locally on an elementary GPU compute block. Then, the full application must be organized around this given building block size. Now, due to phenomena involved in the data transfers between various memory levels across CPUs and GPUs, the optimality of such a scheme is only loosely predictable (as communications tend to overcome in time the complexity of computations). Therefore a mix of (theoretical) analytic approach and (practical) runtime validation is here needed. As we shall illustrate, this occurs at both stage, first at the level of deciding on a given elementary FFT block size, then at the full application level.

\end{abstract}

%A category including the fourth, optional field follows...
%\category{D.2.8}{Software Engineering}{Metrics}[complexity measures, performance measures]


\keywords{OpenCL, GPGPU, FFT, synthetic aperture radar, SIMD, SIMT, Mapping, Algorithm Adaptation}

\section{Introduction}
The Fast Fourier Transform (FFT) Algorithm is one of the top ten algorithms of the 20th century \cite{cipra_best} and it is a basic building block of many signal processing algorithms, including defense systems (warfare radars for example). In such applications one generally needs to compute a number of FFTs iteratively, of large size(s).
There is a plethoric literature on how variants of FFT algorithm (with different number of stages, radixes, etc) may be preferred in general depending on performance features of distinct computing architectures \cite{mirkovic2001automatic,puschel2005spiral,frigo1998fftw,frigo_fast_1999}. But such choices are only tendencial, given that large FFT computations divided into several stages include a lot of data movements to reorder temporary outputs, and that, while computation costs can usually be accurately assessed, those memory transfers are often only loosely predictable.

The case of modern processors such as Intel IvyBridge and Haswell, where CPU is combined with GPU accelerator kernels, adds to this issue's complexity: there exists a given size of register memory available in each GPU computational unit (thereafter called EU, {\em execution unit}), which can be used to define the largest FFT block size that can be computed in a fully local fashion \cite{volkov_fitting_2008}. Now this block could be used as new modular unit for granularity, so that the full-size FFT, and then the whole application, is built around such coarse-grain modular units. But then the communication costs themselves may be prohibitive in that version (beyond being loosely predictable only). So the proposed approach first attempts to compute analytically some sort of a "best version" of the FFT algorithm when regarding the adequation between the FFT computation and the GPU processing and storage power, and then adjust this ideal solution by more practical experiment benchmarking regarding the data transfer and communication efficiency.

\section{Background}
\subsection{FFT Basics}
The FFT factorizes the DFT to reduce the number of computations from $O(N^2)$ to $O(N.log(N))$ for faster evaluation of the discrete Fourier transform (DFT).
The discrete Fourier transform of a signal of N complex samples is given by:

\begin{center}
$ X(k) = \sum\limits_{n=0}^{N-1} x(n).\omega_N^{nk} $ ~~~ with $\omega_N = e^{-2.i.\frac{\pi}{N}}$
\end{center}

The $W_n$ coefficients, commonly known as the twiddle factors, are generally precomputed and stored in memory for reuse. The backward DFT is obtained by changing the sign of the $W_n$ exponent. The FFT algorithm runs multiple stages which are a set of mutiply-add operations named radix or butterfly. For instance, a $2^{12}$ samples radix2 FFT has 12 stages of $2^{11}$ radix2 operations or 6 stages of $2^{10}$ radix4 operations (or 4 stages of $2^9$ operations). If the number of samples does not allow to do exclusively one type of radix operations (for instance one cannot complete a 1024-samples FFT with only radix8 stages), then different radix are used (figure 1) and the implementation is called a mixed-radix FFT.
\begin{figure}[h!]
    \begin{center}
        \includegraphics[width=0.45\textwidth]{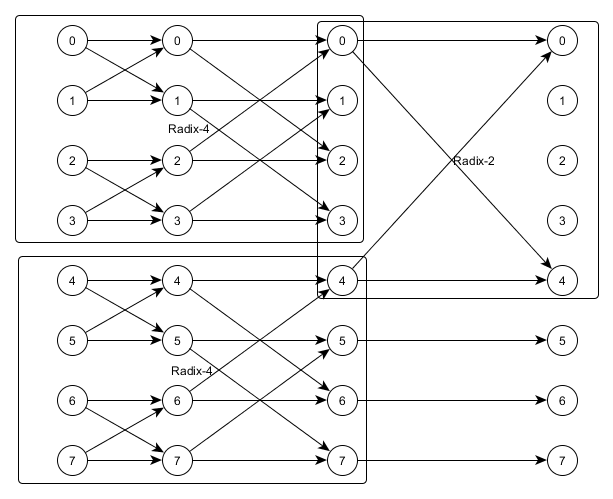}
        \caption{Mixed Radix FFT 8 points}
    \end{center}
\end{figure}

 The mixed-radix implementations are popular because lower radix generally offer poor performance. Most of the signal processing applications which use the FFT perform their operations on a power-of-two number of input samples dataset, with a dataset usually not larger than $2^{12}$. Thus we focused on the mixed-radix implementation.
 
\subsection{Radar application}
\begin{figure}[!h]
\centering
  \includegraphics[width=0.9\linewidth]{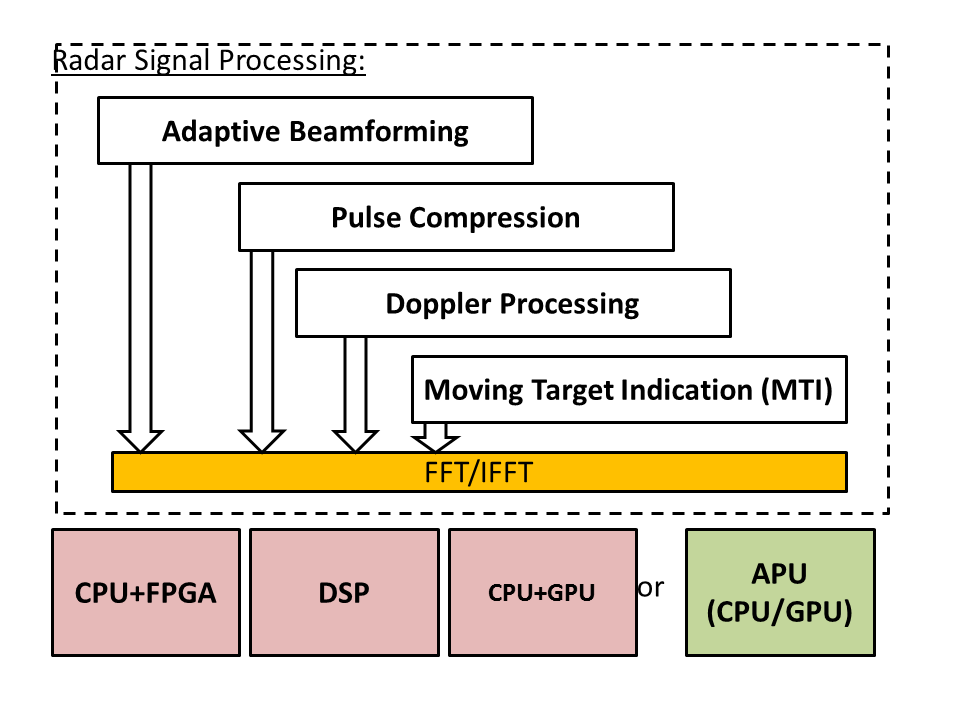}
\caption{Hardware choices for designing an embedded radar system}
\end{figure}
With any military electronics application, balancing size, weight and power (SWaP) presents the biggest challenges to designers. This is especially true for radar systems used on unmanned aerial vehicles (UAV).
For radar applications, GPGPU is now becoming an efficient ingredient to reduce the footprint of the solution when customers want to deploy new radar systems across the world.\\
Exploiting the integrated GPU in the CPU to perform useful radar computations results in the reduction of hardware cost and increases the energy efficiency of the system.\\
SAR(Synthetic aperture radar) is typically mounted on a moving platform such as an aircraft or spacecraft, and it originated as an advanced form of side-looking airborne radar. The distance the SAR device travels over a target creates a large "synthetic" antenna aperture (the "size" of the antenna).

\begin{figure}[!h]
\centering
  \includegraphics[width=1\linewidth]{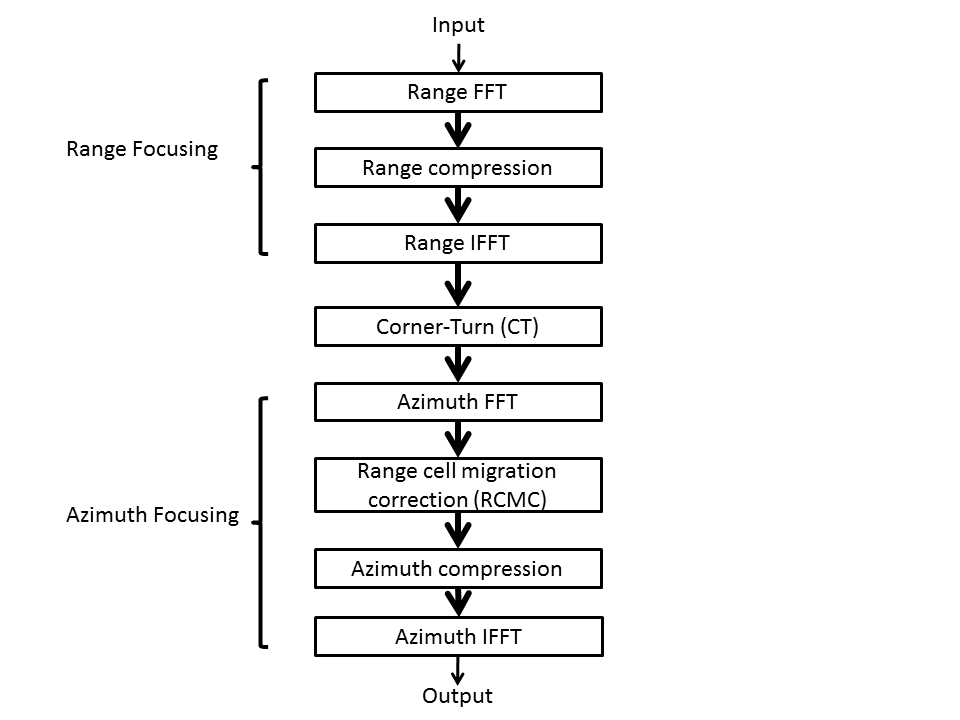}
\caption{synthetic aperture radar range doppler algorithm}
\end{figure}
in this study we will focus on synthetic aperture radar and especially on Range doppler algorithm (RDA)(figure 2).\\

\subsection{Integrated Intel GPGPUs}

The analysis and experiments will be conducted with integrated GPUs. The presence of a CPU and a GPU on the same die opens significant opportunities for parallel algorithms to be accelerated by the GPU. An integrated GPU shares RAM with the CPU (figure 2). This also means some of the cache levels are shared between the CPU and the GPU. In that specific case the L3 cache is shared.
The inner SIMD multithreaded processor (EU: execution unit) architecture is not disclosed. It is not studied in detail but its effects are taken into account in the approach. The number of execution units and their capabilities vary with different GPU implementations\cite{hammarlund2014haswell}. For instance the GPU integrated in Haswell has 40 EUs (HD Graphics 5200GT3e) while the one integrated in the Ivy Bridge CPU has 16 EUs (HD Graphics 4000)(figure 2).
\begin{figure}[!h]
\centering
        \includegraphics[width=0.47\textwidth]{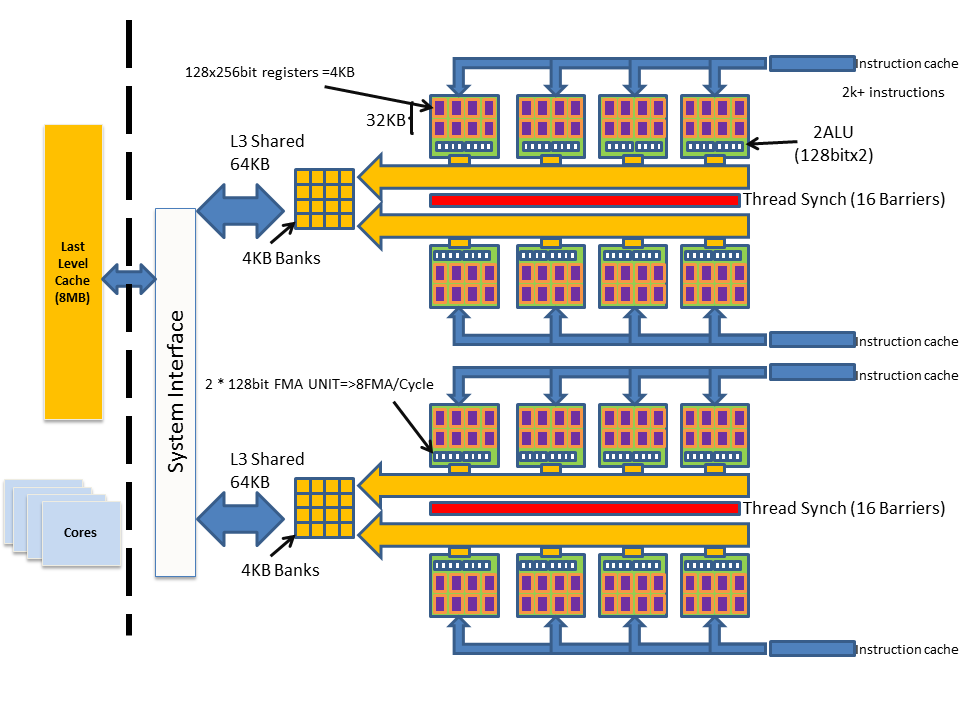}
        \caption{A generic integrated GPU architecture}
\end{figure}
The experiments are conducted with these two integrated CPU/GPU. The third and fourth generation of Intel CPUs (IvyBridge and Haswell) implement the AVX SIMD unit that can provide 8 single precision parallel computations (256bit); maximizing the usage of this unit is not yet efficiently automated by the state of the art compilers, this task is still dedicated to the programmer that can access this unit through intrinsics. Haswell also introduces the fused multiply add instruction and the AVX2.0 SIMD units\cite{hammarlund20134th}.

\section{Approach}

We consider two different layer of optimization, which correspond to two different architecture layers. We will call those two levels of the optimization process respectively the top-level and bottom-level analysis.The bottom-level analysis attempts to optimize the run of one 1024-samples FFT on one computing element (EU).The top-level analysis attempts to optimize the run of a 1024 batch of 1024-samples FFT on the whole system which is a multicore CPU with an integrated multi-EU GPU.\\

In our approach the 1D FFT is tailored to fit the integrated GPU architecture by the means of OpenCL (Open Computing Language)\cite{gaster_heterogeneous_2011}.The OpenCL framework \cite{khronos2008opencl} defines two possible ways to express parallel computations, the first one is the SIMT (Single instruction multiple Threads) programming style that makes an abstraction of the SIMD (Single instruction multiple Data) units, so the programmer is freed from the vectorization issues; the second one is the SIMD vector style.
Our experiments showed that the SIMT fashion is more appropriate for taking benefit of the scalability of GPU with many cores. Open standards are more trustable and OpenCL is now a de facto standard in comparison with Nvidia CUDA\cite{nvidia2011nvidia}.
\begin{figure}[h!]
    \begin{center}
        \includegraphics[width=0.45\textwidth]{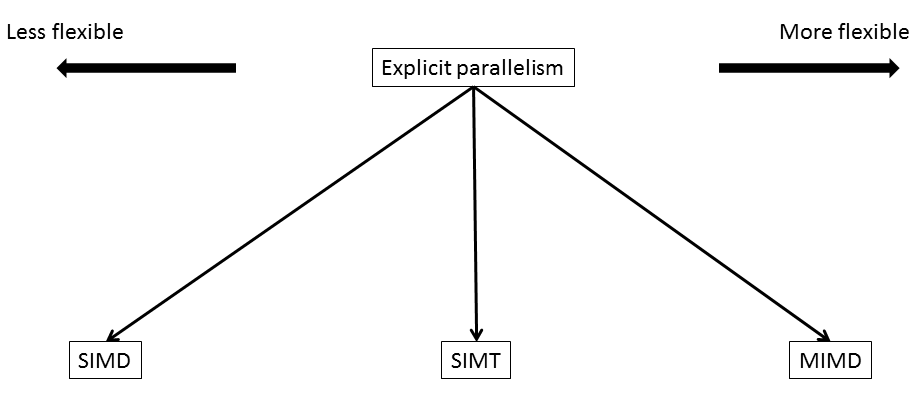}
        \caption{Explicit parallelism and his flexibility}
    \end{center}
\end{figure}

The FFT computation was expressed as a sequence of six FMA operations(multiply-Add), this transformation allows using the full width of the SIMD units within each GPU core (also known as: EU).\\
We have identified three key limiting factors that were used to design our high performance FFT algorithm:
\begin{itemize}
\item Number of registers available on the integrated GPU
\item Size of shared memory 
\item The on-chip interconnect
\end{itemize} 
We have fine-tuned our parallel FFT implementation to maximize the number of floating point operations and to minimize the communications overhead between threads.\\

\subsection{Radar application description}

In order to simplify the understanding we consider a simplified application model (figure 4). This application\cite{skolnik_radar_2008} is massively parallel as for one complete execution the first and last task have to run 1024 times (while the transposition runs only once)(figure 6), and those runs are independent. Those many FFT occurrences are mapped on the Streaming Multiprocessors and the CPU cores. One of the outcomes of this study is to determine how to balance the load between the GPU and the CPU.\\

Moreover some parameters of the FFT block implementation can be tuned as to reach better performance thus we study this block more precisely on the computing elements (EU and CPU core) which will be used to execute it.

\begin{figure}[h!]
\centering
  \includegraphics[width=.35\linewidth]{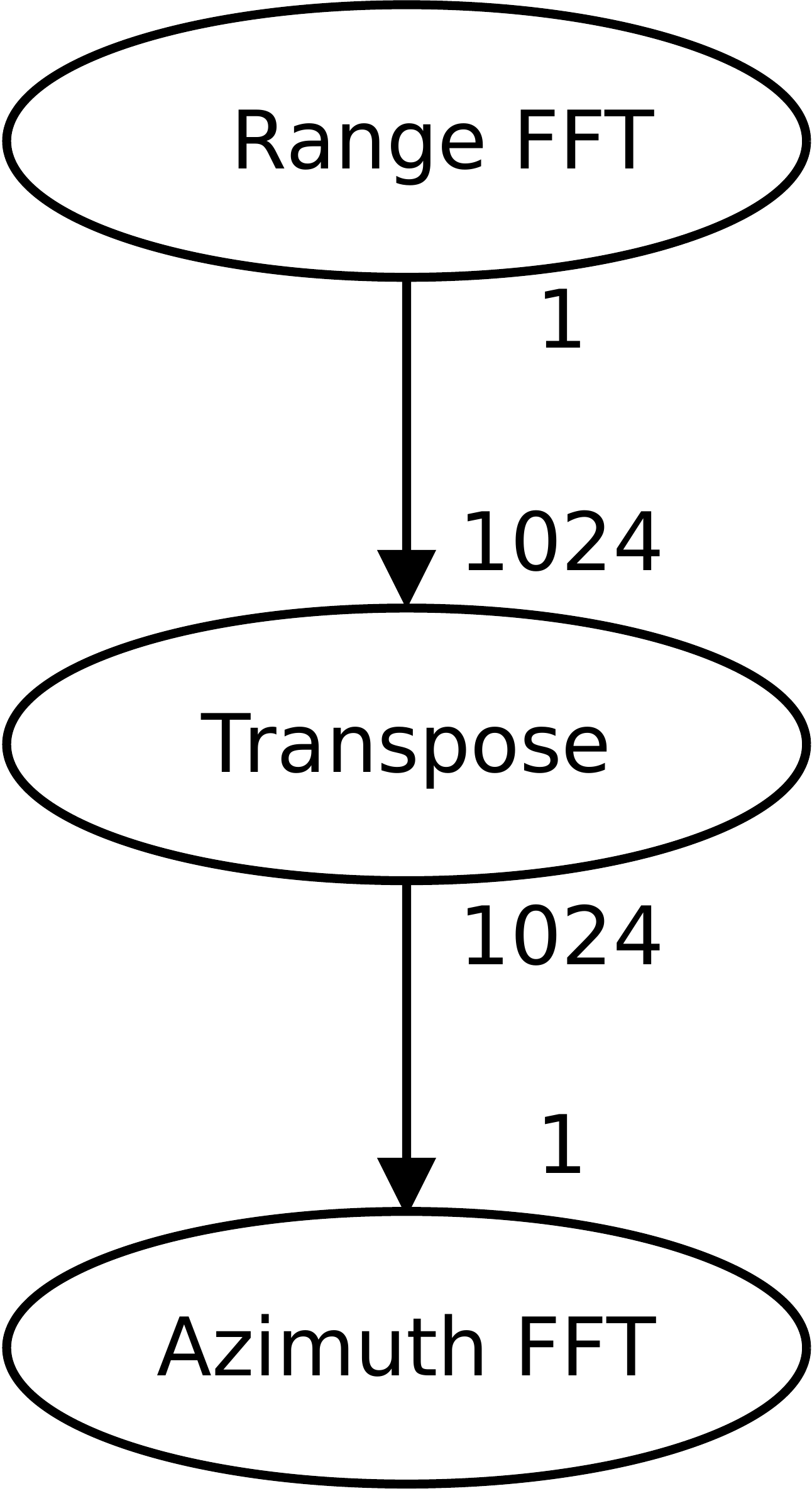}
  \label{fig:radar}
  \includegraphics[width=0.6\linewidth]{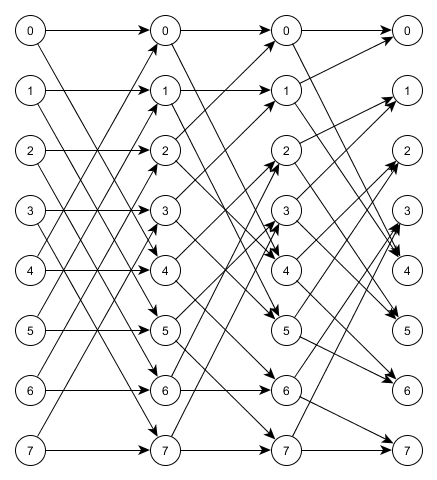}
  \label{fig:butterflies}
\caption{The radar application and the inner FFT block description (8-samples radix2)}
\end{figure}

\subsection{Bottom level analysis}
\subsubsection{Insights:}
Butterflies can be clustered into higher order radix. In order to minimize the stress on the memory hierarchy, we consider that the highest possible radix is the one which input data fits in the private memory of the processing elements. Some simple experiments show that different clusterings will cause different performance gain or loss on the GPU. With $FFT_{list~of~radixes}$ the performance of the FFT implementation when the $list~of~radixes$ is executed, we get on Ivy Bridge:\\
\begin{center}
$\frac{FFT_{2,2,2,2,2,2,2,2,2,2}}{FFT_{8,8,8,2}} = 1.26 $ 
\end{center}
With the insight that $FFT_{8,8,8,2}$ is the optimal implementation, this shows that the naive implementation is much slower than the assumed optimal implementation. We believe however that there exists another list of radixes which performs better than $FFT_{8,8,8,2}$.

The number of different mixed-radix FFT implementations can be very large even for a small set of radixes and ordinary FFT sizes. In our representation it corresponds to the number of distinct paths between the source and the sink nodes. The number of stages that we need to experiment in this approach is $3.(N-1)$ with $N=log_2(nsamples)$. For instance with a 4096-samples 2,4,8 mixed-radix FFT running on a 3.60 GHz Intel Xeon Pentium (according to the performances from the FFTW benchmark page), testing the whole set of combinations would last about 12 seconds while testing only our subset of benchmarks would last much less than one second (only 33 stages runs). We can thus determine a performance model which can help us to choose one particular $FFT$ in order to get performance gains.

\begin{figure}[h!]
    \centering
\begin{tabular} {|l|p{2cm}|p{2,5cm}|}
    \hline
    FFT size & Number of mixed-radix & Number of experiment stage runs\\
    \hline
    16 &	7 &	9\\
    \hline
    32 &	13	& 12\\
    \hline
    64 &	24 &	15\\
    \hline
    128	& 44 &	18\\
    \hline
    256 & 81 &	21\\
    \hline
    512	& 149 &	24\\
    \hline
    1024 &	247 &	27\\
    \hline
    2048 &	504 &	30\\
    \hline
    4096 &	927 &	33\\
    \hline
    8192 &	1705 &	36\\
    \hline
    16384 &	3136 &	39\\
    \hline
\end{tabular}
\caption{Number of mixed-radix possible implementations vs FFT size}
\end{figure}
We identified a bandwidth bottleneck as a limiting factor; this was measured by a memory benchmark between the CPU and the integrated GPU.\\
Our experiments showed that 5GBytes/s is the maximum measured bandwidth on the \textit{Intel IvyBridge (IVB)} ; this gives us also a hint about the maximum achievable FFT GFlops in this architecture. 
\textit{Let B be the memory bandwidth (GB/s)} , \textit{N the FFT size} and \(T_{max}\)\textit{ the maximum throughput(GFlops)}
\begin{eqnarray}
T_{max}= \frac{5.N.\log_2(N)*B}{2*4*N}
\end{eqnarray}
The theoretical maximum performance(1) is 32GFlops for a 1K complex FFT.\\
This was also confirmed by running our previous CPU-GPU bandwidth test on the fourth generation Intel GPU integrated in the \textit{Haswell (HSW)} CPU; the maximum bandwidth being 10GBytes/s. This gives us an upper bound of the GPU maximum performance throughput of 62GFlops.\\
All these theoretical bounds were verified by our implementation of the FFT (figure 10) on these two GPU architectures. \\

\subsubsection{FFT performance model:}
One 1024-samples FFT execution is a successive number of radixes of potentially different orders. Looking at the FFT stages costs shows that there is no straightforward relationship between the index of the stage and its cost. Under the hypothesis that the radix stage performance depend only on the index of the stage (and not on which stages have been run before or will run after) we thus need to benchmark the possible radix runs and find out which combination yields the best performance. The hypothesis is admissible because two stages of the same FFT cannot run in parallel for algorithmic reasons (a synchronization barrier exists between two stages).

\begin{figure}[!h]
    \includegraphics[scale=0.42]{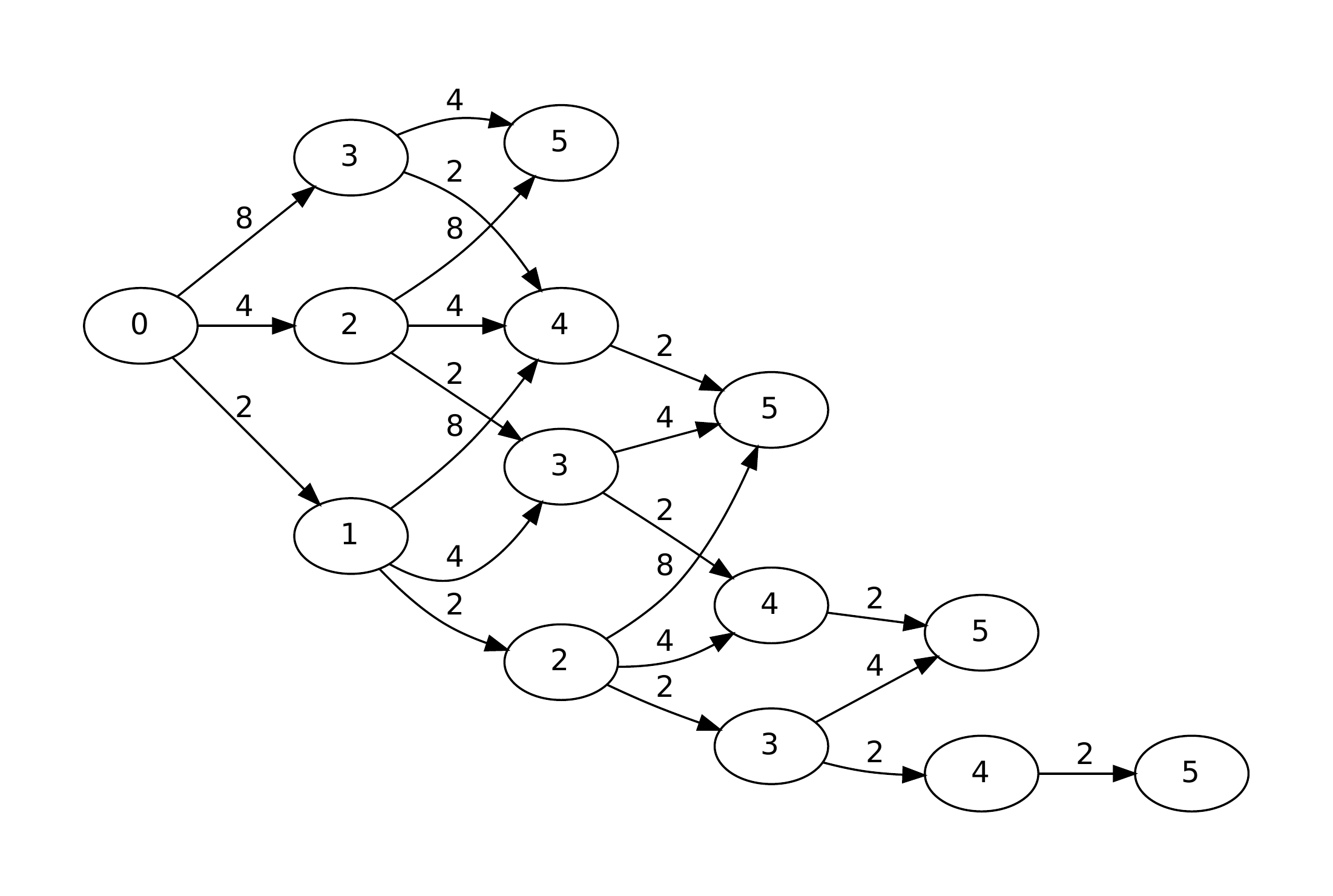}
    \caption{Admissible mixed-radix space for a $2^{5}$ fft}
    \label{fig:fft5}
\end{figure}

The combination of possible radix schedules can be described with a digraph where every path from the input node to any output node is a valid execution. Every edge is weighted with the benchmarked cost and the minimum path from source to sink is the optimal mixed-radix implementation. The figure \ref{fig:fft5} provides the state space of the 32-samples FFT for radixes 2, 4, and 8. For clarity the edge is annotated with the executed radix instead of its cost, and the nodes are annotated with the current stage. The number of possible mixed-radix implementations is the number of distinct paths from the node indexed 0 to the nodes indexed 5. If the edges are weighted with the cost of the executed radix, the shortest path in this list of paths is the optimal mixed-radix 32-samples FFT implementation on Ivy bridge.

In order to determine this path the user needs to benchmark:
\begin{itemize}
    \item radix 2 starting at stage 0, 1, ... 9
    \item radix 4 starting at stage 0, 1, ... 8
    \item ... until the highest possible radix (radix 8 in our experiments)
\end{itemize}

The weights on graph (figure 8) are the selected radixes, to optimise our FFT code we used a shortest path algorithm (in our case: Dijkstra's algorithm). Our benchmarks will provide us precise values depending on the underlying hardware.

\subsubsection{CPU mixed-radix implementation:}
The CPU cores FFT implementation is straightforward because the SIMD units will clearly yield the best performance compared to the regular floating point units, and the size of the SIMD units is fixed and known. The most efficient radix is thus the one which fills the SIMD unit completely. Thus the FFT implementation on the CPU is potentially not the same than the one on the GPU. We implemented an optimized version of FFT using Intel intrinsics to take benefit from the AVX2 units, our implementation shows very close performances  to ones provided by the Intel IPP library. We get 21GFlops and ipp provides 22GFlops in the same conditions.

\subsection{Other FFT sizes}
The approach can be adapted to other FFT sizes. The largest fft basic bloc is fixed in our approach but thanks to the recursive aspect of the FFT, larger FFTs can still be realized following the same methodology, with the same outcomes. Because the Streaming Multiprocessors private memories can hold only 1024 complex samples, and if no other computing resource can hold 4096 samples the stages 11 and 12 need to be split (for instance according to \cite{walker1989portable}) which means the application dataflow graph needs to include this split and merge, which will increase the need for message passing. Applications which require smaller FFTs can follow the same approach.

\begin{figure}[h!]
    \begin{center}
    \includegraphics[width=0.5\textwidth]{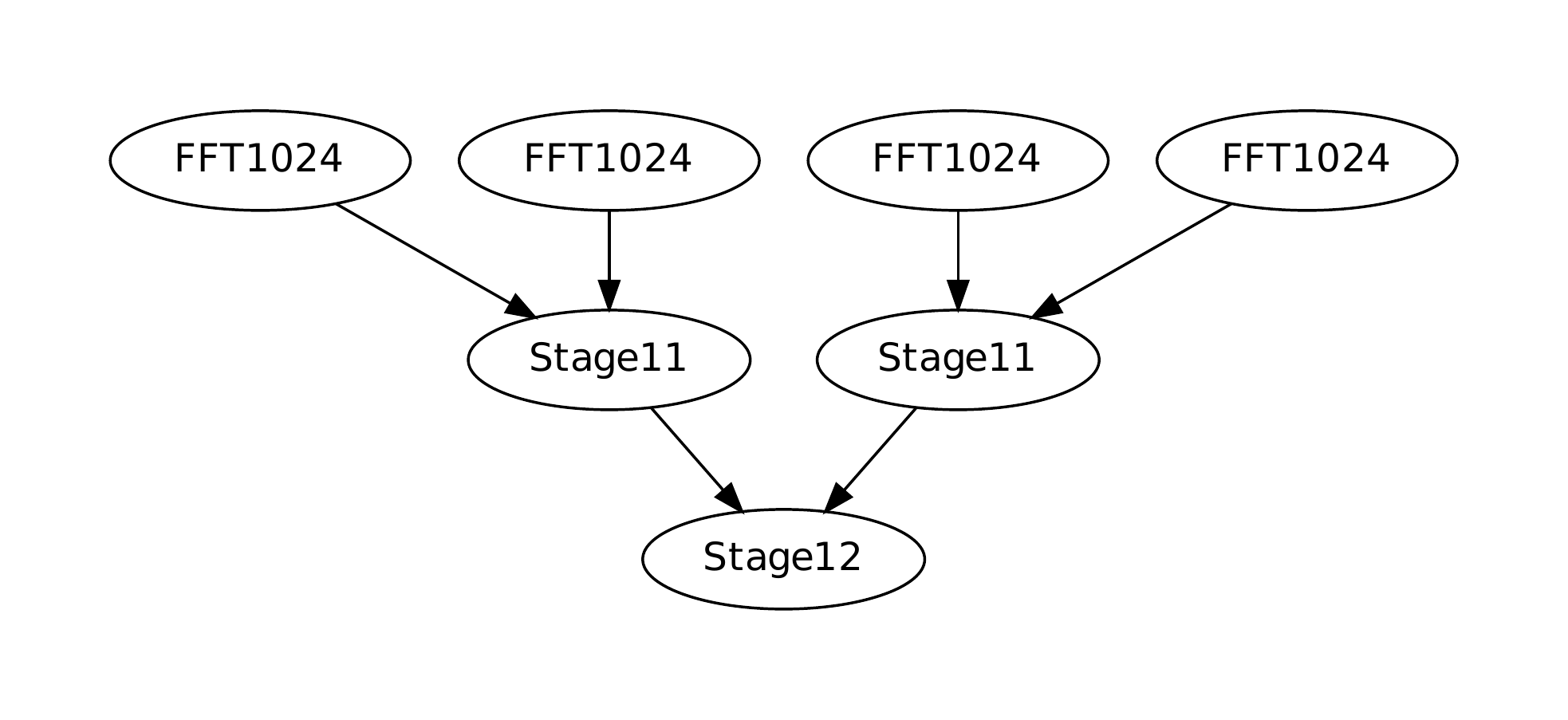}
    \caption{4096-samples FFT described as a combination of 1024-samples FFT}
    \end{center}
\end{figure}

\subsection{Top level analysis}
We intentionally map no more than one FFT to one EU: The EU can hold in its private memory the whole FFT sample set (and no more than one). Once the cost of a FFT is determined, we assume it does not depend on what the other processing elements of the system are doing. This hypothesis is assumed correct because we restricted the sizes of input data such that it fits into the private memories of the processing elements, thus the processing elements will not load much the shared memory hierarchy apart at the first stage (reading) and at the last stage (writing) of the FFT. The performance model for the study of the radar application relies on a dataflow description, annotated with sizes on edges and cost on nodes.
This behavior (Read/Compute/Write) allows us to predict better how the memory hierarchy is being loaded. Thus we assume that the task (nodes on figure \ref{fig:radar}) semantics are:
\begin{itemize}
    \item Read on all its inputs
    \item Compute
    \item Write on all its outputs
\end{itemize}

The experiments show that the performance can vary depending on how the FFT blocks are clustered together on the CPU and on the GPU. Thus there is an optimal repartition ratio which depends on the number of FFTs that need to be executed. The transposition block is not studied but it is well known that a matrix transpose on a large matrix can be split into smaller block \cite{ruetsch2009optimizing} thus given the size of the basic block, the same analysis could be conducted.

\section{Results}
We compare in (figure 9) the performance of our FFT on integrated GPUs (red and blue) to the Intel IPP FFT\cite{taylor2007optimizing} on the CPU (green).
\begin{figure}[h!]
\centering
  \includegraphics[width=1\linewidth]{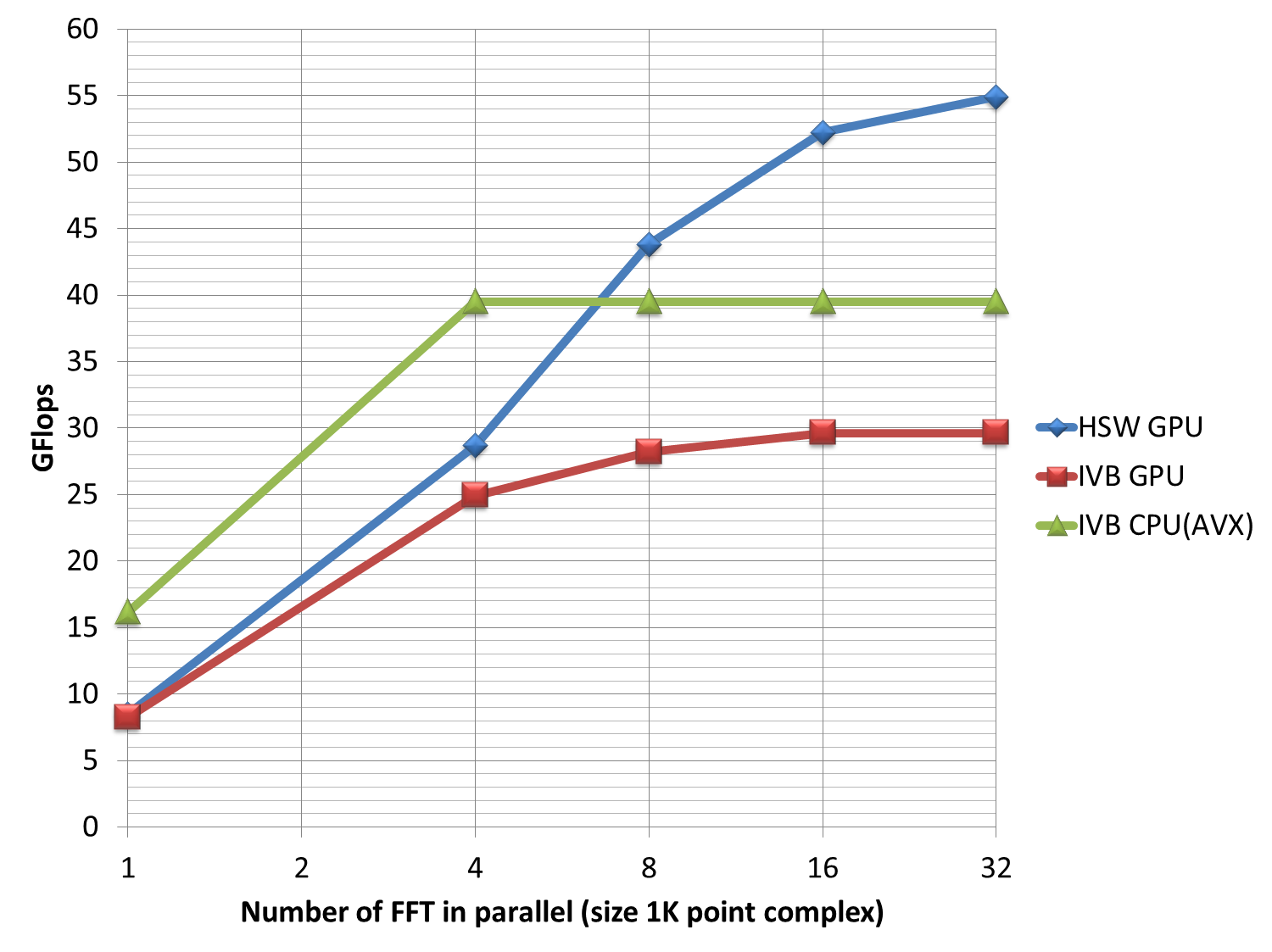}
\caption{Performance scaling of the FFT computation}
\end{figure}
The memory bandwidth gives us a precious indication about our optimization performance, the number of available registers gives us a valuable hint about which algorithm to choose and what is the granularity of our FFT implementation (Radix-8 in our case) and finally the shared memory space defines the number of threads that must be used to process an N size FFT.
\begin{figure}[!h]
\centering
  \includegraphics[width=1\linewidth]{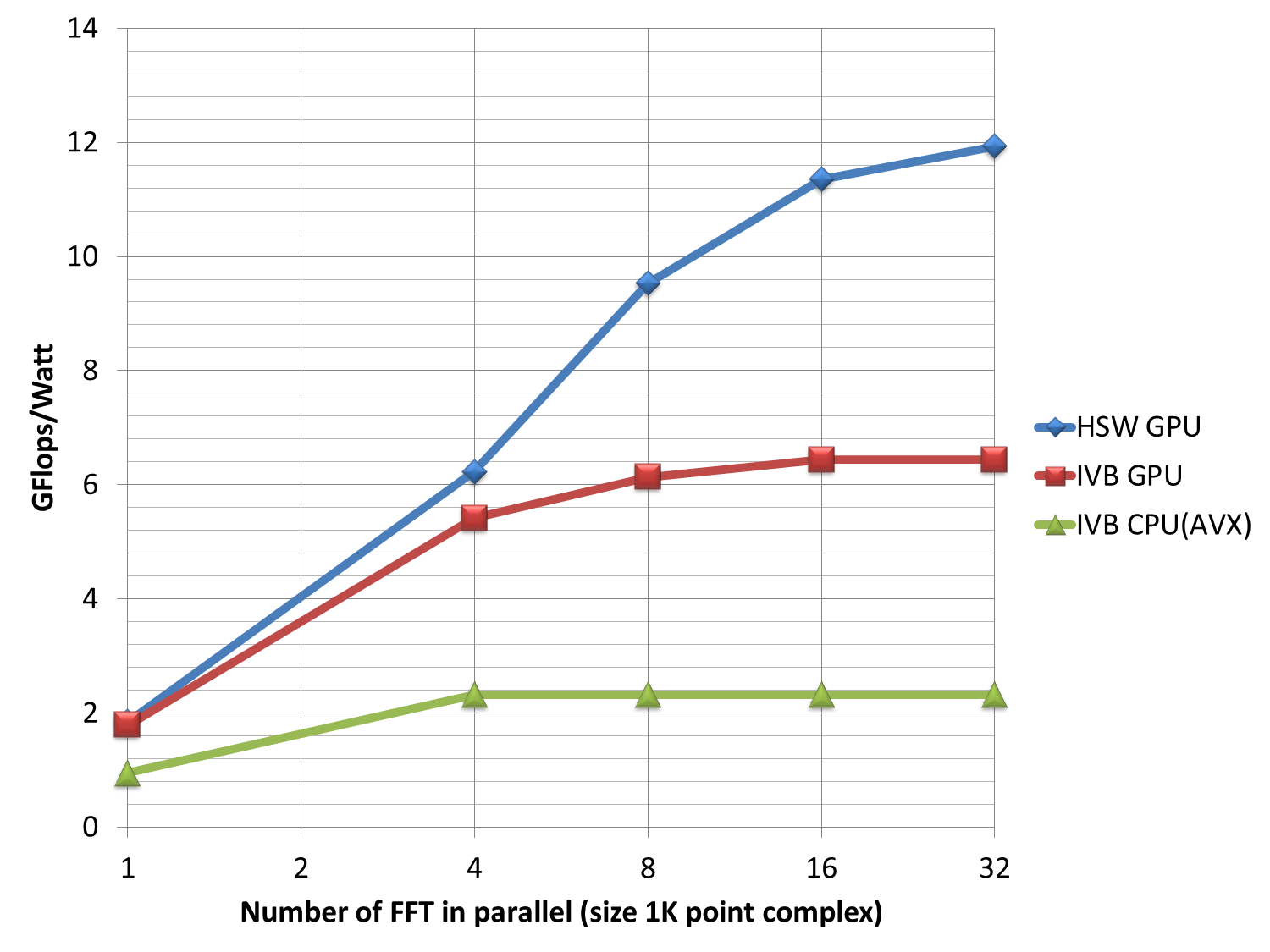}
\caption{Energy efficiency expressed in GFlops per Watt}
\end{figure}

We measured the energy consumption during the FFT computation. The measures in (figure 10) show a low energy footprint of the integrated GPU (4.6W) compared to the CPU. We also noticed that computing the FFT by the integrated GPU entirely frees the CPU for other tasks such as handling the data stream from the sensors and executing the communications software stack.

\begin{figure}[h!]
\begin{minipage}{.5\textwidth}
  \centering
    \begin{tabular}{|l|l|l|l|}
        \hline
        stage & radix 2 & radix 4 & radix 8 \\
        \hline
        0 & 1600 & 3100 & 4135\\
        \hline
        1 & 2430 & 3660 & 4830\\
        \hline
        2 & 2600 & 3600 & 5520\\
        \hline
        3 & 2658 & 4002 & 5988\\
        \hline
        4 & 2560 & 4213 & 6480\\
        \hline
        5 & 2790 & 3910 & 7320\\
        \hline
        6 & 2600 & 4632 & 7896\\
        \hline
        7 & 2889 & 4510 & 7887\\
        \hline
        8 & 3512 & 5030 & X\\
        \hline
        9 & 3913 & X    & X\\
        \hline
    \end{tabular}
    \caption{Intel Ivy Bridge GPU benchmarks subset}
    \label{tab:ivbresults}
\end{minipage}
\end{figure}

\begin{figure}[h!]
\begin{minipage}{.5\textwidth}
    \centering
    \begin{tabular}{|l|l|l|l|}
        \hline
        stage & radix 2 & radix 4 & radix 8 \\
        \hline
        0&1514 & 2813 & 3927\\
        \hline
        1&2295 & 3463 & 4569\\
        \hline
        2&2460 & 3407 & 5223\\
        \hline
        3&2513 & 3785 & 5666\\
        \hline
        4&2427 & 3985 & 6129\\
        \hline
        5&2640 & 3695 & 6930\\
        \hline
        6&2460 & 4385 & 7472\\
        \hline
        7&2737 & 4266 & 7445\\
        \hline
        8&3321 & 4240 & X\\
        \hline
        9&3705 & X    & X\\
        \hline
    \end{tabular}
    \caption{Intel Haswell GPU benchmarks subset}
    \label{tab:hwresults}
\end{minipage}
\end{figure}

Provided the results of figure \ref{tab:ivbresults} we are able to determine that $FFT_{4,8,8,4}$ is the optimal mixed-radix 1024-samples FFT implementation, which shows the naive $FFT_{8,8,8,2}$ was not the optimal one.
The obtained $FFT_{4,8,8,4}$ has 5\% performance gain compared to the $FFT_{8,8,8,2}$ and 31\% performance gain compared to the naive $FFT_{2,2,2,2,2,2,2,2,2,2}$ implementation.
\begin{center}
$\frac{FFT_{8,8,8,2}}{FFT_{4,8,8,4}} = 1.05 $~~~
$\frac{FFT_{2,2,2,2,2,2,2,2,2,2}}{FFT_{4,8,8,4}} = 1.31 $
\end{center}

The same experiments are conducted on the Haswell GPU (figure 13). $FFT_{4,8,8,4}$ has also been found as the optimal mixed-radix implementation with the following performance gains:
\begin{center}
    $\frac{FFT_{8,8,8,2}}{FFT_{4,8,8,4}} = 1.08 $~~~
    $\frac{FFT_{2,2,2,2,2,2,2,2,2,2}}{FFT_{4,8,8,4}} = 1.36 $
\end{center}

The chosen CPU implementation is $FFT_{8,8,8,2}$. We use the SIMD units of the CPU. Yet higher radixes could provide better performance as the fftw-wisdom tool suggests\cite{frigo_design_2005}. The heterogeneous evaluation is conducted on Haswell.

It might be valuable to consider that the CPU can be busy with other tasks, most typically handling TCP/IP communications. Thus it must not be saturated with signal processing work. The ratio can be adapted depending on the desired CPU load.
\begin{figure}[h!]
\centering
  \includegraphics[width=1\linewidth]{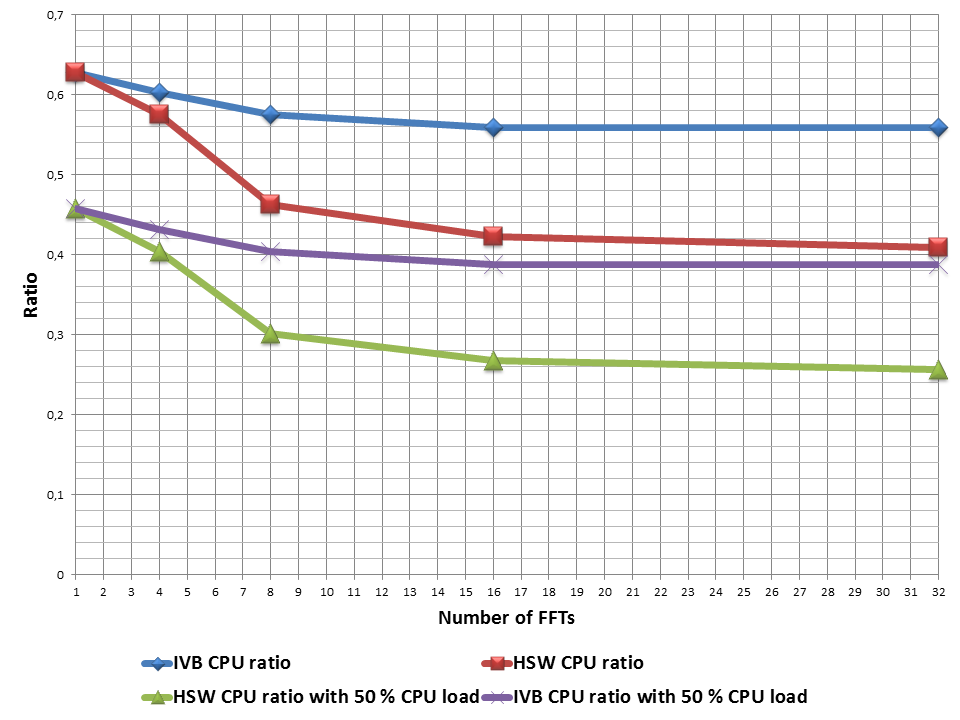}
\caption{The obtained performance for CPU and GPU FFT implementations and the maximum performance CPU ratio}
\end{figure}
Given:
\begin{itemize}
    \item $S_{CPU}$  the number of FFT which run on the CPU 
    \item $S_{GPU}$  the number of FFT which run on the GPU
    \item $S$ the 1024 FFT batch (the total amount of work to be processed)
    \item $P_{CPU}$  the performance obtained on the CPU
    \item $P_{GPU}$  the performance obtained on the GPU 
    \item $CPU_{ratio}\in [0,1]$ the normalized amount of FFTs which runs on the CPU.
\end{itemize}
\begin{center}
$ \frac{S_{CPU}}{P_{CPU}} = \frac{S_{GPU}}{P_{GPU}}$\\
\end{center}
 and 
 \begin{center}
 $S_{CPU}+S_{GPU} = S \iff CPU_{ratio} = \frac{S_{CPU}}{S} = \frac{1}{(\frac{P_{GPU}}{P_{CPU}} + 1)} $
 \end{center}

For the 1024-FFT batch we obtain a performance of 40 GFlops for the CPU and 55 GFlops for the Haswell GPU. Thus the optimal $CPU_{ratio}$ is $42\%$. In order to obtain a $50\%$ load on the CPU, the CPU ratio needs to be set to $27\%$.

\section{Future work}
The present paper investigates how, under a strict specification of GPU EU sizing, one can deduce a theoretically optimal FFT modular brick size, and then build the spatial and temporal organization of a full application by  using such elementary algorithmic component. Because of less predictive behaviors on data transfers and communications, experimental validation may be needed to grasp fast tuning of the whole design. Approaches using Worst-Case Execution Time (WCET) for the data movement operations could unify the latency models, but at the risk of suboptimality, while the choice of which variant of FFT to use may be very sensible to data transfer (either less but larger, or more frequent but smaller).
In the future we would like to consider more involved program shapes for surrounding applications, and other types of parametric algorithms than FFT. The  objective shall remain, to consider further how the interplay between theoretically optimal design campaigns, based on simplified timing assumptions, can be finely re-tuned afterwards by practical experiments, comforting or challenging the theoretical optimality, due to real phenomena hard to predict at model level (such as the relative imprecision of data transfer latency in grey box setting of merged traffics).

\subsection{Related work and optimized mapping}
We provide a brief overview of the related work on optimization techniques that target FFTs on GPUs and CPUs. Van Loan \cite{loan_computational_1992} provides an overview of FFT algorithms and their variants. Frigo and Johnson \cite{frigo_fast_1999} presented FFTW an adaptive library for the efficient computation of FFT of real and complex data of arbitrary dimensions and sizes on many architectures. It employs a two-stage adaptation methodology to adapt to microprocessor architecture and memory hierarchy. At the installation time, the code generator automatically generates highly optimized small DFT code blocks called codelets, At run-time, the pre-generated codelets are assembled in a plan to compute large FFT problems. The space representing various compositions of factorizations and algorithms for a given size FFT is explored to find the best plan of execution. Spiral \cite{puschel2005spiral} is a generator for optimized FFT libraries on CPUs and FPGAs. UHFFT \cite{mirkovic2001automatic} tries to find the best schedule of execution through better understanding of the correlation between the schedules and their performance on modern architectures. The hardware characteristics of GPUs \cite{volkov_fitting_2008} vary widely with newer generation: GPUs now offer better memory hierarchy support, including larger local storage and register file sizes, memory bus widths, etc. Thus it is non-trivial to optimize these algorithms for a distinct range of GPUs.\\
The idea of computing FFTs with 6 FMA operations per butterfly is present in \cite{kaltenberger03fma}. Adjusting the variant of FFT algorithm of that sort to the number of local registers can be found in\cite{volkov_fitting_2008}.\\
While we started this work independently from these sources, our original contribution remains in the study of the subsequent data traffic between CPU, global memory and local registers, and its ability (not quite full) to cope with the full computation bandwidth, while this inability is compensated by the gain in energy spent, a clear winner for embedded computing.\\

\section{Conclusion}
The described approach shows that different modeling levels can be used in order to achieve overall optimization of a signal processing system. The integrated GPGPUs offer a new powerful shared memory vector unit which still offers more mapping and scheduling decisions. Exploring these choices exhaustively can be very time consuming. In our approach we reduce this time-consuming task to a set of simple operations which can help to decide (provided our hypothesis are verified) the optimal solution.
\label{sec_conclu}
We consider our findings  are a step further toward a best performance scaling on modern parallel embedded systems, and also a real opportunity for embedded system designers to make energy efficient systems.\\
Our FFT algorithm can be used in complex and critical applications and a significant performance improvement is expected for no effort.\\
We believe that GPGPUs bringing a 100X speedup and more is a pure myth, and realistic expectations are in general below half of the available computation power, we can clearly argue that 40\% of the peak performance can be reached with a relatively small adaptation effort of the algorithm. This is a valid expectation with the existing architectures, but with the new unified heterogeneous memory hUMA (heterogeneous Uniform Memory Access) that is used in the HSA (Heterogeneous System Architecture)\cite{6487618}, we anticipate productivity increase for the programmer and significant performance gain.

%\end{document}  % This is where a 'short' article might terminate

%ACKNOWLEDGMENTS are optional
\section{Acknowledgments}
The authors would like to thank Mark Martin and Benoit Sansoni for helpful discussions. We
also thank the anonymous reviewers for their valuable comments.
\bibliographystyle{abbrv}
\bibliography{sigproc}  % sigproc.bib is the name of the Bibliography in this case
% ACM needs 'a single self-contained file'!
%\balancecolumns
%\appendix
%Appendix A

\end{document}